\documentclass[mmnp]{edpsmath}

\usepackage[pdftex]{graphicx}
\usepackage{pgfplots}
\graphicspath{{pics/}}
\usepackage{float}

\usepackage{hyperref}
\usepackage[noabbrev]{cleveref}
\numberwithin{equation}{section}
\usepackage{todonotes}
\usepackage[style=numeric, sorting=none, backend=biber, maxnames=999, minnames=3, giveninits=true]{biblatex}
\DeclareNameAlias{default}{given-family} 
\DeclareFieldFormat[article]{title}{#1} 
\DeclareFieldFormat[article]{journaltitle}{\textit{#1}} 
\DeclareFieldFormat[article]{volume}{\textbf{#1}} 
\DeclareFieldFormat{pages}{#1}

\DefineBibliographyStrings{english}{
  andothers = {\textit{et al.,}},
  and = { and }
}

\renewbibmacro{in:}{} 
\AtEveryBibitem{\clearfield{number}}

\usepackage{xcolor}

\begin{document}
\title{Mathematical model of blood coagulation during endovenous laser therapy}

\author{Anna A. Andreeva}\address{Moscow Institute of Physics and Technology (National Research University), 9 Institutskiy per., Dolgoprudny, Moscow Region 141701, Russia.}
\author{Konstantin A. Klochkov}\sameaddress{1}
\author{Alexey I. Lobanov}\sameaddress{1}

\runningtitle{Mathematical model of blood coagulation during endovenous laser therapy}
\runningauthors{A. A. Andreeva \emph{ET~AL.\/}}

\begin{abstract}
Endovenous laser therapy (ELT) as a minimally invasive procedure for ablation of large superficial veins, nevertheless, can cause complications of thrombotic nature. In this regard, the study of the main patterns of thrombus formation during ELT and modelling of endovenous heat-induced thrombosis (EHIT) is relevant. Based on the assumption of diffusion limiting of biochemical processes occurring during the coagulation of blood, by recalculating the reaction rates according to the Stokes–Einstein equation, a simple point model of blood coagulation during ELT was built in this paper. As a result of the use of this model, it was demonstrated that blood heating entails an increase in the rate of thrombin production, a decrease in the time for achieving the peak of its concentration by 5--6 times with its almost constant amplitude. Heating leads to the rapid formation of fibrin clusters and the appearance of a fibrin-polymer network with a smaller cell size. The quantitative dependence on the selected rheological model was also shown. All the data necessary for using the model are given in this article for reproducibility.
\end{abstract}



\maketitle

\section{Introduction}

Endovenous laser therapy (ELT) has become a minimally invasive procedure for the ablation of large superficial veins in patients with varicose veins and venous insufficiency \cite{mozes2005extension}. Despite the safety and effectiveness of the method, thrombotic complications such as deep vein thrombosis or pulmonary embolism may occur. There are studies of such thrombotic complications based on clinical data, for example, \cite{lin2012vein}.

It is becoming relevant to study the main patterns of thrombus formation during ELT. In this case, mathematical modelling of endovenous heat-induced thrombosis (EHIT) processes should play a significant role.

Laser-induced thrombosis is a complex process involving factors of various natures. When considering the problem, it is necessary to take into account complex physical phenomena (absorption of laser radiation by a substance, heating and heat loss due to thermal conductivity, phase transition during cavitation). Hemodynamic conditions (blood flow in the vessel, including filtration flows inside the forming thrombi) and the rheological properties of the medium play an important role. Cavitation in liquids during laser heating of blood has been studied, for example, in \cite{chudnovskii2024features,guzev2023numerical,compton2013hydrodynamic}.

Rheological models of blood depend, for instance, on hematocrit \cite{quemada1978rheology}. Hematocrit, in turn, depends on temperature \cite{abela1985vitro}. Careful consideration of biochemical phenomena is necessary, including blood coagulation reactions, in particular, thrombin generation and subsequent fibrin polymerization. 
When studying such complex "multiphysical" processes with a noticeable lack of experimental and clinical data and a large number of unknown parameters, it is possible to formulate separate hypotheses and test their validity at the level of mathematical modelling. To evaluate qualitative effects and test basic hypotheses, greatly simplified mathematical models are sufficient. As certain assumptions are accepted or rejected, mathematical models must become more complex. Ideally, it will be possible to talk about patient-oriented mathematical models. 

We will construct a mathematical model of laser-induced thrombus formation in the venous tree based on the following assumptions:
\begin{enumerate}
\item Laser radiation is completely absorbed in the blood volume, rapidly raising its temperature locally. With rapid laser heating, the radiation does not reach the vessel walls and does not destroy the endothelial cells;
\item The laser heating procedure is quite long compared to the thermal relaxation times, but short enough that the vascular endothelium does not heat up significantly due to thermal conductivity;
\item The processes of protein denaturation in blood plasma, triggered by heating, are slower than the processes of coagulation and fibrin polymerization. For example, according to \cite{barinov2018thermal}, the time of thermal denaturation of fibrin at 56 $^\circ C$ in the experiment is about 10 minutes;
\item It is possible to estimate the percentage of destroyed cells (platelets, erythrocytes) in the area of laser radiation absorption.
\end{enumerate}
   
Based on the assumptions made, a simplified mathematical model of laser-induced thrombus formation is constructed. We will focus on the issues of thrombin generation, fibrin polymerization, and fibrin thrombus growth characteristic of the venous tree. We will ignore the spatial aspects associated with convective heat transfer and coagulation factors in the vessel. At this stage, we will also not consider phase transitions. We will construct a point (zero-dimensional) mathematical model of the processes of thrombin generation and fibrin polymerization in response to a sharp change in the temperature of the environment.

\section{Mathematical model}

\subsection{Initiation of the coagulation cascade, dynamics of factor XII}

The blood coagulation cascade is conventionally divided into four distinct phases: initiation, amplification, propagation, and stabilization \cite{furie2008mechanisms, versteeg2013new}. Consideration of the initiation phase usually begins with the expression of tissue factor (TF) at the site of vascular damage, which leads to the activation of coagulation factors IX and X by factor VIIa \cite{camerer2000tissue}. The TF--VIIa complex formed at the site of damage links the intrinsic and extrinsic pathways of coagulation. Due to the assumption made about the physics of the process, there is no need to consider the initiation of thrombin production by contact between the active form of coagulation factor VII (normally present in the bloodstream) and tissue factor TF (extrinsic coagulation pathway). Such contact is only possible in the event of endothelial damage. 

In the case under consideration, thrombin generation in the initiation phase is only possible via the intrinsic pathway. When cells (platelets, erythrocytes) are destroyed in the bloodstream, there are a significant number of procoagulant membrane fragments \cite{abela1985vitro}. They can cause the activation of coagulation factor XII. Other causes of factor activation include the presence of neutrophils, partially coagulated proteins, and heparin secreted by mast cells in the bloodstream \cite{didiasova2018factor}. Activated factor XII acts as a catalyst in the activation of factor XI.

Typical solutions for the dynamics of the variables of the mathematical model \cite{panteleev2006spatial} are as follows \cite{panteleev2020differential}. If coagulation reactions begin with the binding of factor VIIa to TF and the formation of an extrinsic tenase complex, then the lag period (the time interval between the formation of the extrinsic tenase and the onset of a sharp increase in thrombin) is relatively short. In this case, fibrin begins to form as soon as thrombin appears in the system. Fibrin production proceeds very quickly and ends long before the thrombin concentration reaches its maximum. With an increase in the initial concentration of tissue factor, the lag period decreases, the width of the thrombin peak also decreases, and the maximum thrombin concentration increases significantly.

The rate of conversion of fibrinogen to fibrin and, as a result, the formation of fibrin clusters and its polymerization will be determined by the dynamics of thrombin production. Note that many reactions can occur not only in solution but also on the surface of membranes, in particular platelet membranes \cite{susree2017mathematical}.

When the system is activated by factor XIa, the processes proceed faster than when activated by extrinsic tenase. Since thrombin concentrations of 10--30 nM \cite{mann2003all} are sufficient for the almost complete cleavage of fibrinogen and the formation of a fibrin polymer gel, it is sufficient to consider only the initiation and amplification phases in the mathematical modelling of the initial stage of laser-induced thrombus formation.

Factor XII (Hageman factor) is activated first. Let us introduce the notation $[XII](0)$ for the initial concentration of Hageman factor in the blood, $[XIIa]$ for the current concentration of the active form of the factor, and $[K]$ for the concentration of kallikrein. Then, for the dynamics of Hageman factor activation, we can write
\begin{equation} 
\frac{d[XIIa]}{dt}=k_{\text{eff}}^{\text{sum}}([XII](0)-[XIIa])+k_{\text{eff}}^{\text{XII,XIIa}}([XII](0)-[XIIa])[XIIa]+k_{\text{eff}}^{\text{XII,K}}([XII](0)-[XIIa])[K]-h^{\text{XIIa}}[XIIa], \label{eq:XIIa}
\end{equation}

\begin{equation}
\frac{d[K]}{dt}=k_{\text{eff}}^{\text{PK,XIIa}}([XII](0)-[XIIa])[XIIa]-h^{\text{K}}[K]. \label{eq:K}
\end{equation}

In \cref{eq:XIIa}, the first term includes the effect of factor activation by all available activators with a certain total "effective" constant. The second term describes the autocatalysis of factor XII, and the third term describes the catalysis by plasma kallikrein. Hageman factor is inhibited by serpins, in particular, ATIII \cite{davoine2020factor}. We account for leakage due to inhibition by the fourth term in \cref{eq:XIIa} with a certain effective inhibition constant.

The \cref{eq:K} describes the dynamics of plasma kallikrein. Reactions of factor XII with high molecular weight kininogens are neglected, since these reactions mainly occur on the surface of the vascular wall \cite{terentyeva2015kinetics,maas2018coagulation}.

The initial concentration of factor XII in human blood is normally $[XII](0) = 375~\text{nM}$ \cite{mailer2022update}. The molecular weight of factor Hageman is 80 kDa.
 
Unfortunately, the authors are unaware of any estimates of the kinetic constants for this stage. Therefore, model values have been selected that ensure activation of factor XII. The values of the model constants are given in \cref{tab:1}.

\begin{table}[H]
\caption{The values of kinetic parameters of the \crefrange{eq:XIIa}{eq:K}. All of them were selected to provide activation of factor XII.}\label{tab:1}
\smallskip
\center
\begin{tabular}{lll}
\hline
Name & Value & Units  \\ \hline
$k^{sum}_{eff}$ & 0.000267 & $\mathrm{min^{-1}}$ \\
$k^{XII,XIIa}_{eff}$ & 0.000267 & $\mathrm{nM^{-1}min^{-1}}$ \\
$k^{XII,K}_{eff}$ & 0.0000267 & $\mathrm{nM^{-1}min^{-1}}$ \\
$h^{XIIa}$ & 0.001 & $\mathrm{min^{-1}}$ \\
$k^{PK,XIIa}_{eff}$ & 0.0000267 & $\mathrm{nM^{-1}min^{-1}}$ \\
$h^K$ & 0.005 & $\mathrm{min^{-1}}$ \\
\hline
\end{tabular}
\end{table}

\subsection{Start of thrombin production and amplification phase}

A reduced system based on the model \cite{susree2017mathematical} is used to describe thrombin production, taking into account the analysis of the initial stage. In addition, two equations are included in the system for a physiologically correct description of the initiation stage: for the dynamics of inactive and active forms of factor XI \cite{andreeva2022using}.

Inhibition by antithrombin III (ATIII) in mathematical models can be represented as constant leaks, since the level of ATIII is such that it does not have time to be consumed in reactions.

Thrombin $[IIa]$ is activated (converted from prothrombin $[II]$) under the action of factor Xa and inhibited by antithrombin III:
\begin{equation}
\frac{d[IIa]}{dt}=k_2^{\text{T}}[Xa]([II](0)-[IIa])-h_2[ATIII][IIa]. \label{eq:IIa}
\end{equation}

Here, the following symbols are used: $[Xa]$ is the concentration of the active form of coagulation factor X, $[II](0)$ is the initial concentration of prothrombin in the system.

Note that in this equation, some effective value of the inhibition rate constant is used in the numerical simulation. This is because thrombin is inhibited only in the plasma volume, and the transition to the membrane form leads to the deposition of a certain amount of thrombin and, as a result, a decrease in the inhibition rate.

Factor XI is activated by thrombin and Hageman factor. For the concentration of inactive and active forms of the factor, we will use the equations
\begin{equation}
\frac{d[XI]}{dt}=-\frac{k_{11}^{\text{cat}}[XI][IIa]}{K_{11M}+[Fg]}-k_{\text{eff}}^{\text{XIIa,XI}}[XI][XIIa], \label{eq:XI}
\end{equation}

\begin{equation}
\frac{d[XIa]}{dt}=\frac{k_{11}^{\text{cat}}[XI][IIa]}{K_{11M}+[Fg]}+k_{\text{eff}}^{\text{XIIa,XI}}[XI][XIIa]-h_{11}[ATIII][XIa]. \label{eq:XIa}
\end{equation}

Here, the symbols $[XI]$ and $[XIa]$ denote the concentrations of the inactive and active forms of factor XI, respectively, and $[Fg]$ denotes the concentration of fibrinogen.

Factor IX is activated by factor XIa and inhibited by antithrombin III.
\begin{equation}
\frac{d[IX]}{dt}=-\frac{k_{9,11}^{\text{cat}}[IX][XIa]}{K_{\text{9,11M}}+[IX]}, \label{eq:IX}
\end{equation}

\begin{equation}
\frac{d[IXa]}{dt}=\frac{k_{9,11}^{\text{cat}}[IX][XIa]}{K_{\text{9,11M}}+[IX]}-h_9[ATIII][IXa]. \label{eq:IXa}
\end{equation}

Here, the symbols $[IX]$ and $[IXa]$ are used to denote the concentrations of the inactive and active forms of factor IX.

Factor X is activated by both extrinsic tenase and intrinsic tenase. For the process under consideration, the formation of extrinsic tenase is neglected. Intrinsic tenase is a complex of factors VIIIa and IXa formed on platelet membranes. The rate of intrinsic tenase formation is high, and its concentration is considered proportional to the concentration of factor VIII. This concentration plays the role of a system parameter. For extrinsic tenase, we assume that its concentration is quasi-stationary (the derivative with respect to time is equal to 0). Then we obtain
\begin{equation}
\frac{d[X]}{dt}=-\frac{k_{10}^{\text{cat}}[X][IXa]}{K_{\text{10M}}+[X]}\frac{k_{\text{ten}}^+}{k_{\text{ten}}^-}[VIII], \label{eq:X}
\end{equation}

\begin{equation}
\frac{d[Xa]}{dt}=\frac{k_{10}^{\text{cat}}[X][IXa]}{K_{\text{10M}}+[X]}\frac{k_{\text{ten}}^+}{k_{\text{ten}}^-}[VIII]-h_{10}[ATIII][Xa]. \label{eq:Xa}
\end{equation}

The model constants (kinetic constants and Michaelis constants) for \crefrange{eq:IIa}{eq:Xa} are given in \cref{tab:2}.

\begin{table}
\caption{The values of kinetic parameters for thrombin generation. The rates of formation and decay of the tenase complex are given under the assumption that factor VIII is fully activated.}\label{tab:2}
\smallskip
\center
\begin{tabular}{lllc}
\hline
Name & Value & Units & References \\ \hline
$k^T_2$ & 0.00045 & $\mathrm{nM^{-1}min^{-1}}$ & \cite{hockin2002model} \\
$h_2[\rm ATIII]$ & 4.08 & $\mathrm{min^{-1}}$ & Estimation \\
$k^{cat}_{9,11}$ & 11 & $\mathrm{min^{-1}}$ & \cite{oliver1999thrombin} \\
$K_{9,11M}$ & 60 & $\mathrm{nM}$ & \cite{oliver1999thrombin} \\
$h_9[\rm ATIII]$ & 47.6 & $\mathrm{min^{-1}}$ & Estimation \\
$k^{10}_{cat}$ & 500 & $\mathrm{min^{-1}}$ & \cite{mann1990surface} \\
$K_{10M}$ & 63 & $\mathrm{nM}$ & \cite{mann1990surface} \\
$\frac{k^+_{ten}}{k^-_{ten}}[{\rm VIII}]$ & 1.4 & -- & \cite{hockin2002model,susree2017mathematical} \\
$h_{10}[\rm ATIII]$ & 0.68 & $\mathrm{min^{-1}}$ & Estimation \\
$k^{11}_{cat}$ & 54.6 & $\mathrm{min^{-1}}$ & \cite{oliver1999thrombin} \\
$K_{11M}$ & 50 & $\mathrm{nM}$ & \cite{oliver1999thrombin} \\
$k_{eff}^{XIIa,XI}$ & 0.05 & $\mathrm{nM^{-1}min^{-1}}$ & Estimation \\
$h_{11}[\rm ATIII]$ & 3.4 & $\mathrm{min^{-1}}$ & Estimation \\
\hline
\end{tabular}
\end{table}

\subsection{Fibrin polymerization}

Existing mathematical models of polymerization attempt to describe the polymerization process in isolation from fibrinogen cleavage and fibrin monomer production. They usually assume that fibrin in the system was formed instantaneously and consider the Cauchy problem either for the concentrations of various oligomers or for the distribution moments (oligomers, clusters, protofibrils, gel, \etc.).

We assume that the fibrin polymerization process is quite fast. This is consistent with experimental data \cite{weisel2017fibrin,chernysh2008dynamic} that fibrin polymer gel forms approximately 70 seconds after a lag period of 3 minutes. The main stages of the process are as follows. When thrombin appears in the system, it causes the cleavage of fibrinogen and the formation of fibrin monomer. The characteristic mass of a fibrinogen molecule is 700 kDa, and that of a fibrin monomer is 360 kDa \cite{weisel2017fibrin}. We consider further processes to be diffusion-controlled due to the large size of the molecules.

Due to knob-hole interactions \cite{weisel2017fibrin,mosesson2005fibrinogen}, fibrin monomers combine into oligomers of various lengths and clusters of different configurations \cite{weisel2017fibrin,hunziker1990new}. Neglecting the details of the process, it can be assumed that the increase in cluster size occurs due to attachment to the active group of the fibrin monomer. The attachment of the monomer can also lead to the formation of branches.
\begin{equation}
\frac{d[Fg]}{dt}=-\frac{k_{\text{cat}}^{\text{Fg,IIa}}[Fg][IIa]}{K_M^{\text{Fg,IIa}}+[Fg]}, \label{eq:Fg}
\end{equation}

\begin{equation}
\frac{d[Fn]}{dt}=\frac{k_{\text{cat}}^{\text{Fg,IIa}}[Fg][IIa]}{K_M^{\text{Fg,IIa}}+[Fg]}-(1+\kappa)k_l(2[Fn]+[A])[Fn], \label{eq:Fn}
\end{equation}

\begin{equation}
\frac{d[A]}{dt}=\kappa k_l[A][Fn]+(2+3\kappa)k_l [Fn]^2, \label{eq:A}
\end{equation}

\begin{equation}
\frac{d[C_0]}{dt}=\frac{k_{\text{cat}}^{\text{Fg,IIa}}[Fg][IIa]}{K_M^{\text{Fg,IIa}}+[Fg]}, \label{eq:C0}
\end{equation}

\begin{equation}
C(t)=\frac{(1+\kappa)[A](t)-\kappa([C_0](t)-[Fn](t))}{2+\kappa}, \label{eq:C}
\end{equation}

\begin{equation}
a(t)=\frac{[A](t)}{C(t)}=\frac{(2+\kappa)[A](t)}{(1+\kappa)[A](t)-\kappa([C_0](t)-[Fn](t))}, \label{eq:a}
\end{equation}

\begin{equation}
n(t)=\frac{(2+\kappa)([C_0](t)-[Fn](t))}{(1+\kappa)[A](t)-\kappa([C_0](t)-[Fn](t))}. \label{eq:n}
\end{equation}

The symbols in \crefrange{eq:Fg}{eq:n}: $[Fg]$ is the fibrinogen concentration, $[Fn]$ is the current concentration of fibrin monomers, $[C_0]$ is the total number of fibrin monomers at a given moment, $[A]$ is the concentration of "active groups" (unbound hall domains, D-regions), $C(t)$ is the concentration of forming fibrin clusters (oligomers, protofibrils, fibrils, \etc.), $a(t)$ is the average number of branches per cluster, $n(t)$ is the average number of monomers per cluster. A more detailed description of the fibrin polymerization model, including cluster and branch formation, is given in \cite{panyukov2024scaling}.

The dynamics of fibrinogen $[Fg]$ cleavage under the action of free thrombin is described by \cref{eq:Fg}. \Cref{eq:Fn} describes the dependence of the concentration of monomers on time. The first term in it is responsible for the production of monomers due to the cleavage of fibrinogen to fibrin, the second describes the consumption of fibrin monomers for the formation of clusters. $k_l$ is the rate constant of linear fibrin polymerization. $\kappa$ is the dimensionless branching rate, equal to the ratio of the branching rate to the rate of linear fibrin polymerization.

\Cref{eq:A} describes the dynamics of the concentration of "active groups" (unbound hall domains, D-regions) in the system. Clusters are tree-like structures of polymerized fibrin. Free hall domains are located at the ends of these structures. It is to them that the following monomers can attach.

The system parameters are shown in \cref{tab:3}.

\begin{table}
\caption{The values of kinetic parameters for fibrin polymerization.}\label{tab:3}
\smallskip
\center
\begin{tabular}{lllc}
\hline
Name & Value & Units & References \\ \hline
$k^{Fg,IIa}_{cat}$ & 5040 & $\mathrm{min^{-1}}$ & \cite{higgins1983steady} \\
$K^{Fg,IIa}_M$ & 7200 & $\mathrm{nM}$ & \cite{higgins1983steady} \\
$k_l$ & 0.8 & $\mathrm{nM^{-1}min^{-1}}$ & \cite{lewis1985characterization} \\
$\kappa$ & 0.02 & -- & \cite{panyukov2024scaling} \\
\hline
\end{tabular}
\end{table}

\begin{table}
\caption{Initial concentration of coagulation factors.}\label{tab:4}
\smallskip
\center
\begin{tabular}{llllc}
\hline
Component & Value (nM) & Component & Value (nM) & References \\ \hline
$\rm XII$ & 375 & $\rm XIIa$ & 0 & \cite{mailer2022update} \\
$\rm K$ & 0 & & & \cite{mailer2022update} \\
$\rm II$ & 1400 & ${\rm IIa}$ & 0 & \cite{susree2017mathematical} \\
$\rm VIII$ & 0.7 & ${\rm VIIIa}$ & 0 & \cite{susree2017mathematical} \\
$\rm IX$ & 90 & ${\rm IXa}$ & 0 & \cite{susree2017mathematical} \\
$\rm X$ & 170 & ${\rm Xa}$ & 0 & \cite{susree2017mathematical} \\
$\rm XI$ & 30 & ${\rm XIa}$ & 0 & \cite{panteleev2010task} \\
$\rm Fg$ & 7600 & $\rm Fn$ & 0 & \cite{susree2017mathematical} \\
\hline
\end{tabular}
\end{table}

\subsection{Temperature dependence}

All reaction constants collected in \crefrange{tab:1}{tab:3} were measured or estimated based on data from various coagulation experiments (platelet-rich plasma or donor blood) conducted at a temperature of 37 $^\circ C$. There is data in the literature on the functioning of the coagulation system during hypothermia \cite{wolberg2004systematic, whelihan2014thrombin}. During
hypothermia, the rate of thrombin generation reactions decreases. With laser heating, the temperature can be significantly higher than normal human body temperature.

The most common hypothesis in this case is to use the Arrhenius equation to estimate the reaction rate when the temperature changes. An example of this approach is given in \cite{pfefer2000pulsed,barton2001thermal}. The use of the Arrhenius equation for a complex mathematical model leads to the need to determine a significant number of unknown constants (activation energy values), on which the solution strongly depends.

All substances involved in coagulation reactions have significant sizes and molecular masses of tens or hundreds of kilodaltons (for example, 58 kDa for antithrombin III, 80 kDa for Hageman factor, and 700 kDa for fibrinogen). With such molecule sizes, the reactions will be diffusion-controlled, \ie, their rate is determined by the diffusion of the reacting particles, after which their interaction occurs almost instantaneously. As a rule, such reactions have a low activation energy, in contrast to activation-controlled reactions. For them, the reaction constant is directly proportional to the diffusion coefficient in the liquid.

To estimate the temperature diffusion coefficient in a liquid, we use the Stokes–Einstein equation
\begin{equation}
D=\frac{k_B T}{6\pi\eta r}. \label{eq:einst}
\end{equation}

Here, $D$ is the diffusion coefficient, $k_B$ is Boltzmann constant, $r$ is the effective radius of the molecule, and $\eta$ is the dynamic viscosity of the fluid \cite{einstein1905molekularkinetischen, von1906kinetischen, bromberg2003molecular}.

From \cref{eq:einst} follows the relationship for the diffusion coefficients in a fluid at different temperatures
\begin{equation}
\frac{D(T_1)}{D(T_2)}=\frac{T_1}{T_2}\frac{\eta(T_2)}{\eta(T_1)}. \label{eq:rel}
\end{equation}

\Cref{eq:rel} determines the temperature dependence of the diffusion coefficient. Then, for the rates of diffusion-controlled reactions, we will take \cref{eq:rel} as the temperature dependence:
\begin{equation}
\frac{k(T_1)}{k(T_2)}=\frac{D(T_1)}{D(T_2)}=\frac{T_1}{T_2}\frac{\eta(T_2)}{\eta(T_1)}. \label{eq:krel}
\end{equation}

For Michaelis constants, assuming that the activation energy of the reverse reaction is weakly dependent on temperature, the temperature dependence will be determined by the inverse of \cref{eq:rel}:
\begin{equation}
\frac{K_M(T_1)}{K_M(T_2)}=\frac{k^-(T_1)}{k^+(T_1)}\frac{k^+(T_2)}{k^-(T_2)}=\frac{T_2}{T_1}\frac{\eta(T_1)}{\eta(T_2)}, \label{eq:Krel}
\end{equation}
where $k^+(T)$ is the forward rate constant and $k^-(T)$ is the reverse rate constant at temperature $T$ in Michaelis\-Menten kinetics.

\subsection{Viscosity models}

The specific form of the right-hand side of \cref{eq:rel} is determined by the accepted rheological model of the fluid. In the first-order approximation, blood can be considered a Newtonian fluid, and the temperature dependence can be taken, for example, from the Frenkel equation \cite{frenkel1955kinetic}. Currently, the Quemada model \cite{quemada1978rheology} is considered to be an adequate rheological model for blood. In the Quemada model, the effective viscosity of blood depends on the hematocrit $Ht$, the shear rate (in a more general case, on the second invariant of the strain rate tensor), and two dimensionless parameters
\begin{equation}
\eta = \eta_0 \left(1-\frac{k_0+k_\infty\sqrt{\frac{|I_2|^{1/2}}{\gamma_{cr}}}}{2\left(1 + \sqrt{\frac{|I_2|^{1/2}}{\gamma_{cr}}}\right)}Ht\right)^{-2}. \label{eq:quemada}
\end{equation}

Another dependence of blood viscosity on hematocrit \cite{pries1992blood} includes an unknown exponent and one empirical parameter
\begin{equation}
\eta=\eta_0(1+B((1-Ht)^C-1)). \label{eq:pries}
\end{equation}

In \cite{pries1992blood}, parameters are estimated for different tube diameters. In model calculations, $B = 2.92$, $C = -0.808$. This corresponds to large tube diameters.

The dependence of hematocrit on blood temperature for calculations is based on the extrapolation of experimental data for laser-induced thrombosis \cite{abela1985vitro}. A logistic function was used for the extrapolation
\begin{equation}
Ht=\frac{1}{1+\exp((T-T_0)/k)}, \label{eq:ht}
\end{equation}
where $T_0=34~^\circ C$ and $k=7~^\circ C$.

In addition, a polynomial model of viscosity was considered
\begin{equation}
\eta=\eta_0\exp\left(B/T + CT + DT^2\right), \label{eq:poly}
\end{equation}
where $B=4209$, $C = 4.527\cdot 10^{-2}$ and $D = -3.376\cdot 10^{-5}$ for water according to \cite{viswanath2007viscosity}.

\section{Calculation results}

The qualitative results of calculations of the thrombus growth process are the same for all the rheological models under consideration. For illustration purposes, the results are given using the Quemada rheological model \labelcref{eq:quemada} and the polynomial rheological model \labelcref{eq:poly}. Calculations were also carried out for the viscosity model \labelcref{eq:pries}, but the results were similar to the Quemada model \labelcref{eq:quemada}, and therefore were not presented in the manuscript.

The model parameters for calculations using the Quemada model \labelcref{eq:quemada} are $k_0 = 4$, $k_\infty = 1.6$. These parameters are typical for calculations of the rheological properties of real blood samples \cite{marcinkowska2007comparison}. Since we are interested in a point model, the ratio of the shear rate to the critical rate was taken as a constant equal to 2. This value corresponds to a shear rate of 10--20 $s^{-1}$. Thus, in \cref{eq:quemada}, the value of the numerical factor before the hematocrit is fixed, and the dependence of the hematocrit on temperature was calculated using the given extrapolation formula \labelcref{eq:ht}.

\begin{figure}[h]
 \centering
  \includegraphics[width=\textwidth]{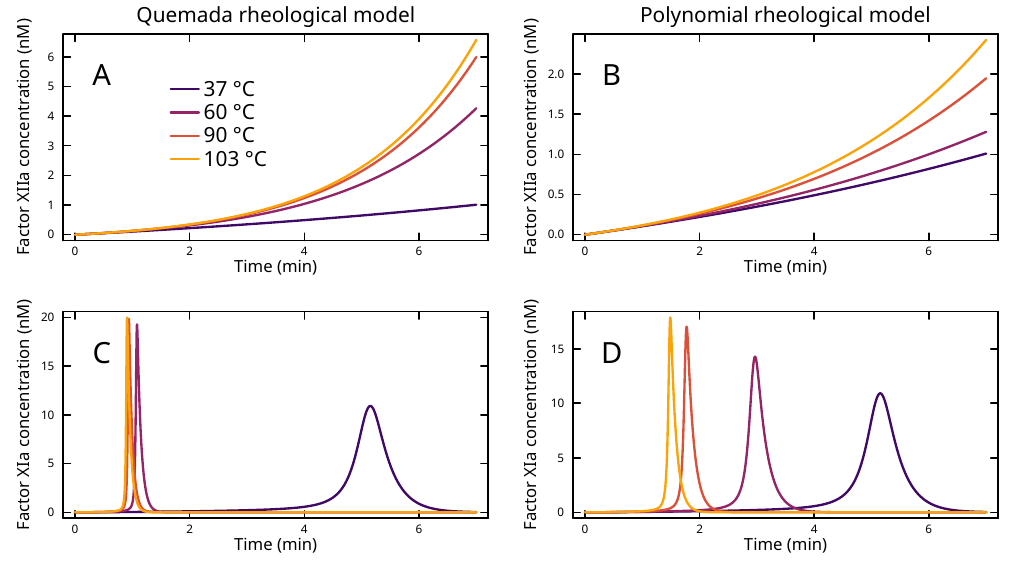}
	\caption{Initiation of the coagulation process. The blood coagulation system is activated via an intrinsic pathway under the influence of factor XIIa. Factor XIIa activates factor XI. Factor XI is activated by thrombin through positive feedback loops.  (A) Dynamics of factor XIIa concentration for the Quemada model at temperatures of 37 $^\circ C$ (blue line), 60 $^\circ C$ (purple line), 90 $^\circ C$ (red line), and 103 $^\circ C$ (orange line). (B) Dynamics of factor XIIa concentration for the polynomial model at temperatures of 37 $^\circ C$ (blue line), 60 $^\circ C$ (purple line), 90 $^\circ C$ (red line), and 103 $^\circ C$ (orange line).  (C) Dynamics of factor XIa concentration for the Quemada model at temperatures of 37 $^\circ C$ (blue line), 60 $^\circ C$ (purple line), 90 $^\circ C$ (red line), and 103 $^\circ C$ (orange line).  (D) Dynamics of factor XIa concentration for the polynomial model at temperatures of 37 $^\circ C$ (blue line), 60 $^\circ C$ (purple line), 90 $^\circ C$ (red line), and 103 $^\circ C$ (orange line).}
\label{pic:1}
\end{figure}

The dynamics of factors XIIa (upper panels) and XIa (lower panels) are shown in \cref{pic:1}. The calculations used the Quemada rheological model \labelcref{eq:quemada} (left panels) and the polynomial model \labelcref{eq:poly} (right panels). The temperature values in the calculations are 37 $^\circ C$, 60 $^\circ C$, 90 $^\circ C$, and 103 $^\circ C$. Despite the fact that when using the Quemada rheological model, factor XII activation occurs faster with increasing temperature, only a small fraction of factor XII (initial concentration value of 375 nM) is activated in the system for both rheological models.

Factor XI is activated by the active form of factor XII. Its further increase is determined by positive feedback loops during thrombin generation. With increasing temperature, factor XI is activated faster, and the peak amplitude increases slightly (lower panels in \cref{pic:1}).

\begin{figure}[h]
 \centering
  \includegraphics[width=\textwidth]{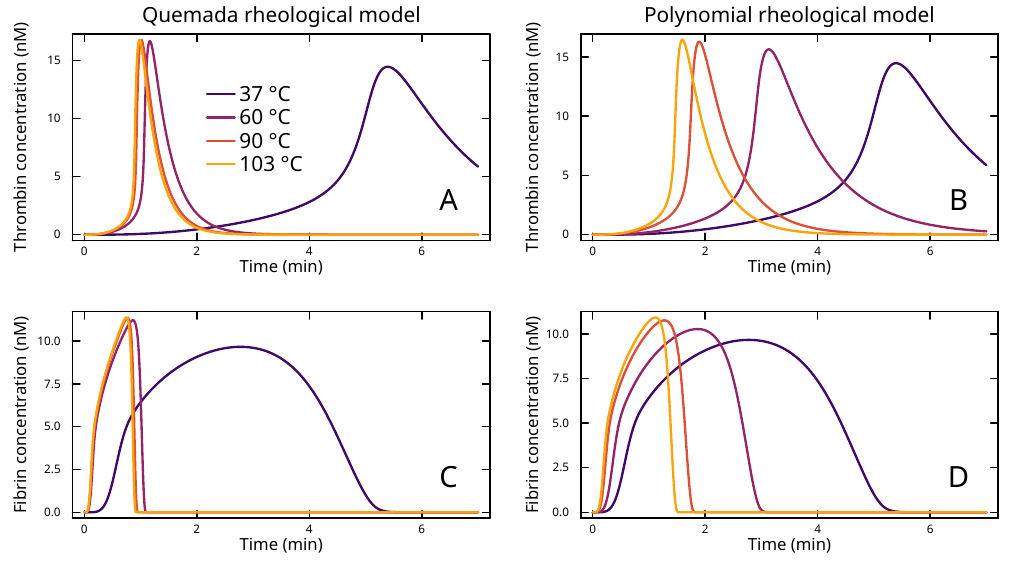}
	\caption{The cascade of blood coagulation factor activation ends with the production of thrombin and the cleavage of fibrinogen molecules into fibrin under the action of thrombin. (A) Dynamics of thrombin concentration for the Quemada model at temperatures of 37 $^\circ C$ (blue line), 60 $^\circ C$ (purple line), 90 $^\circ C$ (red line), and 103 $^\circ C$ (orange line). (B) Thrombin concentration dynamics for the polynomial model at temperatures of 37 $^\circ C$ (blue line), 60 $^\circ C$ (purple line), 90 $^\circ C$ (red line), and 103 $^\circ C$ (orange line). (C) Dynamics of fibrin concentration for the Quemada model  at temperatures of 37 $^\circ C$ (blue line), 60 $^\circ C$ (purple line), 90 $^\circ C$ (red line) and 103 $^\circ C$ (orange line).  (D) Dynamics of fibrin concentration for the polynomial model  at temperatures of 37 $^\circ C$ (blue line), 60 $^\circ C$ (purple line), 90 $^\circ C$ (red line) and 103 $^\circ C$ (orange line).}
\label{pic:2}
\end{figure}

\Cref{pic:2} shows the dynamics of thrombin production (upper panels) and the dynamics of fibrin monomer concentration (lower panels) for the same temperature values and selected rheological models.

The rapid increase in factor XIa concentration leads to an accelerated increase in thrombin concentration. Fibrinogen cleavage and fibrin monomer formation occur during the thrombin concentration increase phase, long before it reaches its maximum value.

According to the model \cite{panyukov2024scaling}, it is the rate of fibrinogen cleavage that determines the structure of the forming fibrin clusters at the initial stage of polymerization, and subsequently the structure of the fibrin polymer network. It can be seen that the production rate changes significantly with temperature. The relatively slow increase in monomer concentration at 37 $^\circ C$ leads to slow polymerization (a long decreasing section after the maximum concentration, lower panels in \cref{pic:2}), while heating leads to rapid cluster formation and rapid formation of the fibrin polymer network.

\begin{figure}[h]
 \centering
  \includegraphics[width=\textwidth]{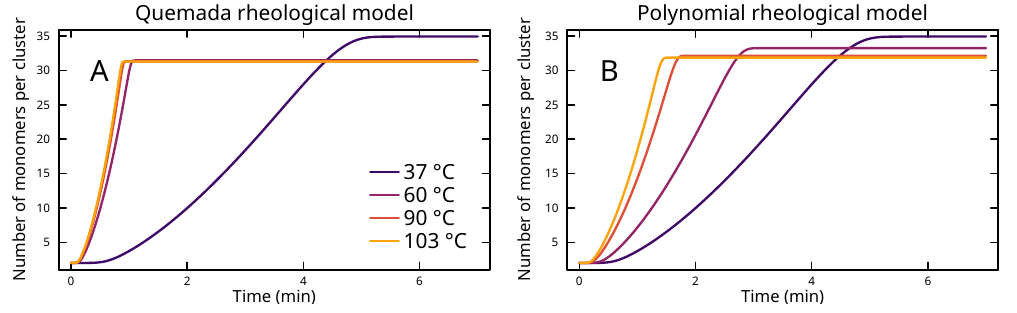}
	\caption{The production of fibrin polymer can be characterized by the number of monomers per cluster. (A) Average number of monomers per cluster for the Quemada model at temperatures of 37 $^\circ C$ (blue line), 60 $^\circ C$ (purple line), 90 $^\circ C$ (red line), and 103 $^\circ C$ (orange line). (B) Average number of monomers per cluster for the polynomial model at temperatures of 37 $^\circ C$ (blue line), 60 $^\circ C$ (purple line), 90 $^\circ C$ (red line), and 103 $^\circ C$ (orange line).}
\label{pic:3}
\end{figure}

Fibrin fibers are characterized by the average number of monomers per cluster, which determines the length of the protofibrils and, as a result, the size of the polymer network cells. As the temperature increases, the average number of monomers per cluster decreases, leading to the formation of shorter fibers. For temperatures of 37 $^\circ C$, 60 $^\circ C$, 90 $^\circ C$, and 103 $^\circ C$ and the selected rheological models, the average number of fibrin monomers in the forming cluster is shown in \cref{pic:3}. As a result, an increase in temperature leads to the formation of a network with a smaller cell size. Note that when using other rheological models, the difference between the average cluster sizes at high temperatures is greater than for the Quemada rheological model. Another characteristic of the cluster is the number of branches. It also decreases slightly with temperature change. Note that this value is also sensitive to the rheological model of the fluid.

\section{Conclusion and Discussion}
The considered simplest mathematical model of thrombus formation during laser heating of blood shows that the hypothesis of diffusion-controlled reactions entails rapid formation of fibrin thrombi. As the temperature increases, polymer networks with smaller cell sizes are formed.

The use of the Stokes–Einstein formula is justified not only for protein molecules whose shape is close to spherical and for which the Stokes formula is valid. During the polymerization of fibrin, when each monomer is added to the polymer chain, the efficiency of its transport is ensured by rotational diffusion. To describe the dependence of the rotational diffusion coefficient, a formula different from \cref{eq:einst} is used. But even for rotational diffusion, the temperature and viscosity of the liquid are linearly included in the numerator and denominator of the ratio, respectively. Thus, for rotational diffusion, it is also possible to use \cref{eq:rel} to modify the reaction rate.

Of interest is the study of the dependence of polymer gel formation on the blood composition. For this, one should rely on detailed mathematical models of coagulation that operate with a large number of factors.

The appearance of bubbles during cavitation leads to the appearance of a phase boundary, which in turn can be a surface on which the generation of thrombin is initiated, the appearance of fibrin with its subsequent polymerization.

The publication considers the simplest point mathematical model of coagulation during rapid heating of blood. In a real blood vessel, an important role is played by the distribution of heat throughout the volume of the vessel, primarily due to convective transfer. Thermal conductivity processes also play a significant role. Taking into account spatial phenomena associated with non-uniform heating can give an idea of the spatial structure of such a thrombus, and, ultimately, serve as a basis for assessing the risks of thrombotic complications.

\begin{acknowledgement}
We are deeply grateful to Vladimir M. Chudnovskii for bringing to our attention the problem that led to this study.
\end{acknowledgement}

\printbibliography
\end{document}